# Reconfigurable Intelligent Surface-Assisted Ambient Backscatter Communications – Experimental Assessment

R. Fara[1,2], D.-T. Phan-Huy[2], P. Ratajczak[2], A.Ourir[3], M. Di Renzo[1] and J. De Rosny[3]
[1]*Université Paris-Saclay, CNRS, CentraleSupélec, Laboratoire des Signaux et Systèmes,*
3 rue Joliot-Curie, 91192, Gif-Sur-Yvette, France
[2]*Orange Labs Networks,* Châtillon & Sophia-Antipolis, France
[3]*ESPCI Paris, PSL University, CNRS, Institut Langevin*, Paris, France
romain.fara@orange.com

***Abstract*—Sixth generation (6G) mobile networks may include new passive technologies, such as ambient backscatter communication or the use of reconfigurable intelligent surfaces, to avoid the emission of waves and the corresponding energy consumption. On the one hand, a reconfigurable intelligent surface improves the network performance by adding electronically controlled reflected paths in the radio propagation channel. On the other hand, in an ambient backscatter system, a device, named tag, communicates towards a reader by backscattering the waves of an ambient source (such as a TV tower). However, the tag's backscattered signal is weak and strongly interfered by the direct signal from the ambient source. In this paper, we propose a new reconfigurable intelligent surface assisted ambient backscatter system. The proposed surface allows to control the reflection of an incident wave coming from the exact source location towards the tag and reader locations (creating hot spots at their locations), thanks to passive reflected beams from a predefined codebook. A common phase-shift can also be applied to the beam. Thanks to these features, we demonstrate experimentally that the performance of ambient backscatter communications can be significantly improved.**

***Keywords—Ambient backscattering; 6G; Internet of things; Reconfigurable intelligent surface; Metasurface.***

## I. INTRODUCTION

The Internet of Things (IoT) is continuously expanding and connecting more people and objects, through wireless links, resulting in a considerable growth of the power consumption due to the emission of radio frequency (RF) waves. Recently, technologies such as ambient backscatter (AmB) communications have been proposed to reduce the power consumption and the emission of RF waves of IoT devices [1]. In an AmB communication system, tags send messages to readers by backscattering RF waves generated by an ambient RF source such as a TV tower, a Wi-Fi access point or a base station from the cellular network. As tags avoid generating additional RF waves, their energy consumption is very low, and solar or RF waves energy harvesting is sufficient to power them [2]. Such tags are thus promising devices for IoT applications [3].

More precisely, in its simplest implementation [1], the tag is a dipole antenna (designed to resonate at the carrier frequency of the RF ambient source) whose two branches are either connected or disconnected thanks to a switch. The tag is illuminated by an ambient RF source. Hence, if the branches are connected, the tag backscatters the ambient waves and the tag is said to be in a backscattering state. If the branches are disconnected, the tag is almost transparent to the ambient waves and the tag is said to be in a transparent state. One state corresponds to bit '0' and the other to bit '1'. A tag sends a binary message simply by switching itself between the two states. A reader nearby receives the ambient waves, and only needs to detect variations in the received power to distinguish the two states and demodulate the binary message (using a non-coherent energy detector) from the tag [1].

Unfortunately, the reliability in terms of bit error rate (BER) of such a simple system is limited by the weakness of the backscattered signal and the simplicity of the non-coherent energy detector. As a first limitation, the tag may be in a deep fade of the ambient signal. In that case, it cannot backscatter any signal and it is undetectable by the reader. To overcome this first limitation, in [4] a massive multiple input multiple output (M-MIMO) antenna is used to beamform the ambient signal and create a hot spot on the tag, to boost the backscattered waves. As a second limitation, the ambient signal interferes with the backscattered signal of the tag and can degrade the BER. Indeed, the authors of [5] show that, even in line-of-sight (LOS), the interferences and the desired signal can combine in such a way that the received signals at the reader side in the two states are close in power and only differ by their phases. In this case, the performance of a non-coherent detector collapses. To overcome this second limitation, the authors of [4] create another additional hot spot, on the reader, this time, with a controlled phase shift, to make sure that both the signals of the source and the tag combine coherently when they arrive at the reader. Such scheme maximises the received contrast (between the two states of the tag) in terms of power.

Currently, another promising technology, named reconfigurable intelligent surface (RIS), is being investigated for future 6th generation (6G) mobile networks [6]. An RIS is an electronically controlled metasurface composed of a large array of passive reflecting elements, which can reflect RF waves in a controlled manner, to allow operators improve the efficiency and the coverage of their network, without requiring any network densification [6,7]. More precisely, an RIS is an array of backscatters, where each element controls electronically and individually the phase-shift (and/or an amplitude and/or a polarization rotation) with which it backscatters an incident wave [8,9,10,11]. By applying a particular set of phase-shifts, the RIS controls the direction of the reflection, i.e. it creates a reflected beam. Phase-shifts can hence be viewed as reflection beamforming weights, similar to active beamforming weights. Such an RIS improves the coverage by creating a reflected beam towards a coverage hole.

In [12], an RIS has been proposed to increase the quality of the signal at the reader location in an AmB communication system. More precisely, it has been shown by means of simulations in rich-scattering environments that an RIS can modify the multipath channel and enhances the system performance. Similarly, in [13] it has been proposed to use an RIS to assist a cognitive backscattering communication system. However, the solutions proposed in [12] and [13] have not been tested and verified experimentally.

In this paper, we propose a new practical RIS-assisted AmB system, to be deployed on an information desk exposing tags for the purpose of reading. The idea of using RISs in AmB systems has been experimentally studied from the information-theroric standpoint in [11], where an RIS communicates a message to a reader, in our system the RIS does not transmit any data but helps existing tags, carrying information, to transmit them to the reader. In our proposed system, the source is the room Wi-Fi access point, it is fixed and in LOS of the desk, and its location is registered and known at the RIS. On the contrary, the exact locations of the tag and reader on the desk are unknown. However, the desk area is split into pre-defined locations. The RIS is close to the desk and has two features. First, the RIS can pick a beam among a pre-defined beam of a codebook. Each beam is designed to reflect waves coming from the exact source location towards one of the pre-defined locations on the desk. Second, the RIS can apply a common phase shift to all its elements. The RIS tests several pairs of beam and common phase-shift. Each time, the reader reports a performance quality feedback (a quantized contrast for instance). The RIS and the reader together search for the beam and common phase-shift pair that maximizes the performance. We expect, with such approach, to improve the BER, with respect to the scenario without RIS, with a similar approach as in [4]: i.e., either by creating a "hot spot" on the tag (and boosting the backscattered wave), or by creating "coherent spots" both on the tag and the reader, in such a way that all waves reflected by the RIS combine coherently with all other waves, when arriving at the reader (to boost the received contrast). Finally, through experiments, we validate our approach and show a significant performance improvement.

The paper is organized as follows: Section II introduces the model of our proposed new system, Section III describes our experimental setup for validation, Section IV illustrates the obtained experimental results and Section V concludes this paper.

*Notations:* bold letters denotes matrices and vectors. We use $|h|$ and $\arg(h)$ to denote the absolute and argument value of the complex number $h$, respectively. $\mathrm{erfc}(\cdot)$ is the complementary error function.

## II. System Model

### *A. Channel modeling*

We consider the communication between one tag and one reader assisted by an RIS in a LOS environment. Our system model, illustrated in Fig. 1, does not take into account the mutual coupling between the RIS elements (cells) [14]. First, the source transmits a signal originally destined to a legacy device. The signal is then reflected by the RIS with $M$ cells and also backscattered by the tag to the reader.

The reader (R) receives the signal from the source (S) through the path $g$ composed of four different paths between S and R. Among them, two paths are controlled by the RIS and two are modulated by the tag (T). The direct path is denoted by "S-R". The multi-hop path that is modulated by the tag is denoted by "S-T-R". The multi-hop path that is reflected by the RIS is denoted by "S-R, RIS". Finally, the path that is reflected by the RIS and is modulated by the tag is denoted by "S-T-R, RIS". Other paths, with more hops, are neglected. Multi-hop paths are composed of single-hop paths: "S-T", "T-R". Table I, defines, for each path, the path-specific notation of the channel coefficient $h$, the amplitude $a$ and the distance $d$. With such notation and definitions, $g$ is given by:

$$g = h^{S-R} + \gamma h^{S-T-R} + h^{S-R,Ris} + \gamma h^{S-T-R,Ris}, \quad (1)$$

where $\gamma$ is a scalar that can take two different values depending on the state of the tag ($\gamma = 1$ if the tag is in the backscattering state and $\gamma = 0$ if the tag is in the transparent state).

TABLE I. NOTATIONS OF THE PROPAGATION CHANNEL COEFFICIENTS FOR ALL PATHS

| Path | Channel coefficient ($h \in \mathbb{C}$) | Phase of the propagation channel coefficient ($\phi \in \mathbb{R}$) | Amplitude ($a \in \mathbb{R}^+$) | Distance ($d \in \mathbb{R}^+$) |
|---|---|---|---|---|
| S-R | $h^{S-R}$ | $\phi^{S-R}$ | $a^{S-R}$ | $d^{S-R}$ |
| S-T | $h^{S-T}$ | $\phi^{S-T}$ | $a^{S-T}$ | $d^{S-T}$ |
| T-R | $h^{T-R}$ | $\phi^{T-R}$ | $a^{T-R}$ | $d^{T-R}$ |
| S-T-R | $h^{S-T-R}$ | $\phi^{S-T-R}$ | $a^{S-T-R}$ | $d^{S-T-R}$ |
| S-R,RIS | $h^{S-R,Ris}$ | $\phi^{S-R,Ris}$ | NA | NA |
| S-T-R,RIS | $h^{S-T-R,Ris}$ | $\phi^{S-T-R,Ris}$ | NA | NA |

Whatever the path without RIS reflection is, the coefficient $h$ of a propagation channel can be approximated by:

$$h = ae^{j\phi}, \quad (2)$$

with $\phi = \frac{2\pi d}{\lambda}$, where $\lambda$ corresponds to the wavelength of the signal. The channel coefficient of a path between a point A and a point B, assisted by the reflection of the RIS, depends on the channel coefficients of the sub-paths between A and each cell of the RIS, and the sub-paths between each cell of the RIS and B.

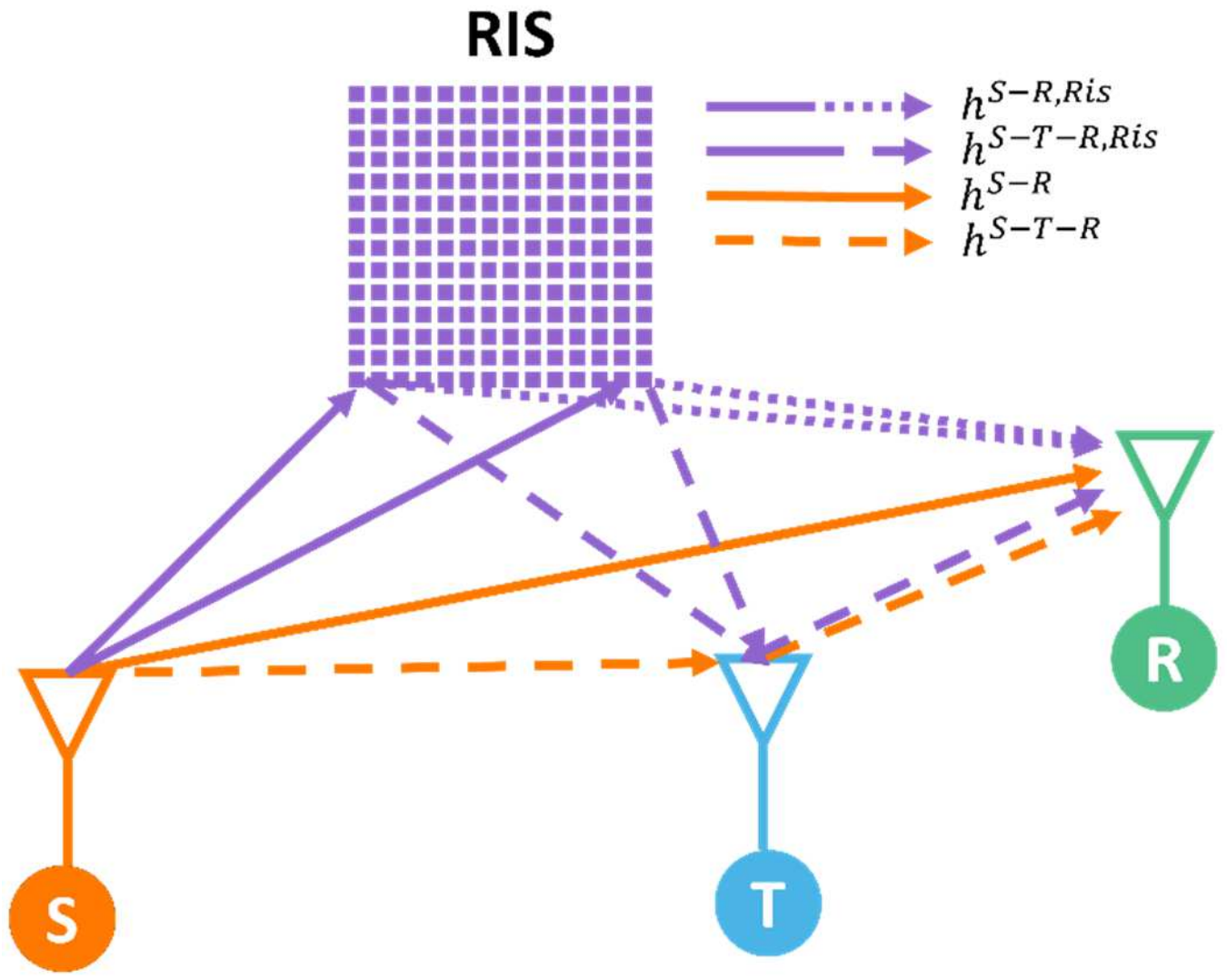

Fig. 1. System model with the links between the source (S), the RIS, the tag (T) and the reader (R).

Hence we define the channel vectors $\mathbf{h}^{Ris-R} \in \mathbb{C}^{M\times 1}$, $\mathbf{h}^{S-Ris} \in \mathbb{C}^{1\times M}$ and $\mathbf{h}^{Ris-T} \in \mathbb{C}^{M\times 1}$ with sub-paths defined in Table II. For each sub-path, the sub-path-specific notation for the channel coefficient $h$, its amplitude $a$ and its phase $\phi$, is provided. Again, whatever the sub-path is, we have: $h = ae^{j\phi}$.

TABLE II. NOTATIONS FOR THE RIS CHANNEL SUB-PATHS

| Sub-Path | Channel coefficient ($\mathbf{h}_m \in \mathbb{C}$) | Phase of the propagation channel coefficient ($\phi \in \mathbb{R}$) | Amplitude ($a \in \mathbb{R}^+$) | Distance ($d \in \mathbb{R}^+$) |
|---|---|---|---|---|
| S-cell m of the RIS | $\mathbf{h}_m^{S-Ris}$ | $\phi_m^{S-Ris}$ | $a_m^{S-Ris}$ | $d_m^{S-Ris}$ |
| cell m of the RIS-R | $\mathbf{h}_m^{Ris-R}$ | $\phi_m^{Ris-R}$ | $a_m^{Ris-R}$ | $d_m^{Ris-R}$ |
| cell m of the RIS-T | $\mathbf{h}_m^{Ris-T}$ | $\phi_m^{Ris-T}$ | $a_m^{Ris-T}$ | $d_m^{Ris-T}$ |

With the notations defined in Table I and Table II, we obtain:

$$h^{S-T-R} = h^{S-T} h^{T-R}, \quad (3\text{-a})$$
$$h^{S-R,Ris} = \mathbf{h}^{S-Ris}\mathbf{u}\mathbf{h}^{Ris-R}, \quad (3\text{-b})$$
$$h^{S-T-R,Ris} = \mathbf{h}^{S-Ris}\mathbf{u}\mathbf{h}^{Ris-T} h^{T-R}, \quad (3\text{-c})$$

where $\mathbf{u}$ is the matrix of the reflection beamforming phase-shift coefficients applied to the cells of the RIS. In this paper, we consider an RIS that can only phase shifts the signal (without amplitude modulation). Therefore, $\mathbf{u}$ is a diagonal matrix, of size $M$, of unitary complex coefficients.

### *B. Energy detector : performance metrics*

On the reader side, we consider an energy detector to detect the tag. The performance of such detector depends on the amplitude of the signal contrast received by the reader $\Delta g$ between the two states of the tag, i.e. $\Delta g = \left| |g_{\gamma=1}| - |g_{\gamma=0}| \right|$. The signal level received by the reader for each state of the tag ($\gamma = 1$ or $\gamma = 0$) can be determined with the equation (1) and the amplitude of the contrast can be expressed, with the RIS $\Delta g^{Ris}$ and without the RIS $\Delta g^{ref}$, as follows:

$$\Delta g^{ref} = \left| |h^{S-R} + h^{S-T-R}| - |h^{S-R}| \right|, \quad (4\text{-a})$$
$$\Delta g^{Ris} = \left| |h^{S-R} + h^{S-T-R} + h^{S-R,Ris} + h^{S-T-R,Ris}| - |h^{S-R} + h^{S-R,Ris}| \right|. \quad (4\text{-b})$$

In these expressions, only the terms $h^{S-R,Ris}$and $h^{S-T-R,Ris}$ are controllable by the RIS.

Considering an energy detector used by the reader to decode the tag message, the performance is given, as in [5] by:

$$BER = 0.5 \text{ erfc}\left(\frac{\Delta g}{\sigma}\right), \quad (5)$$

where $\sigma$ is the square root of the noise average power.

### *C. Codebook-based reflection beamforming*

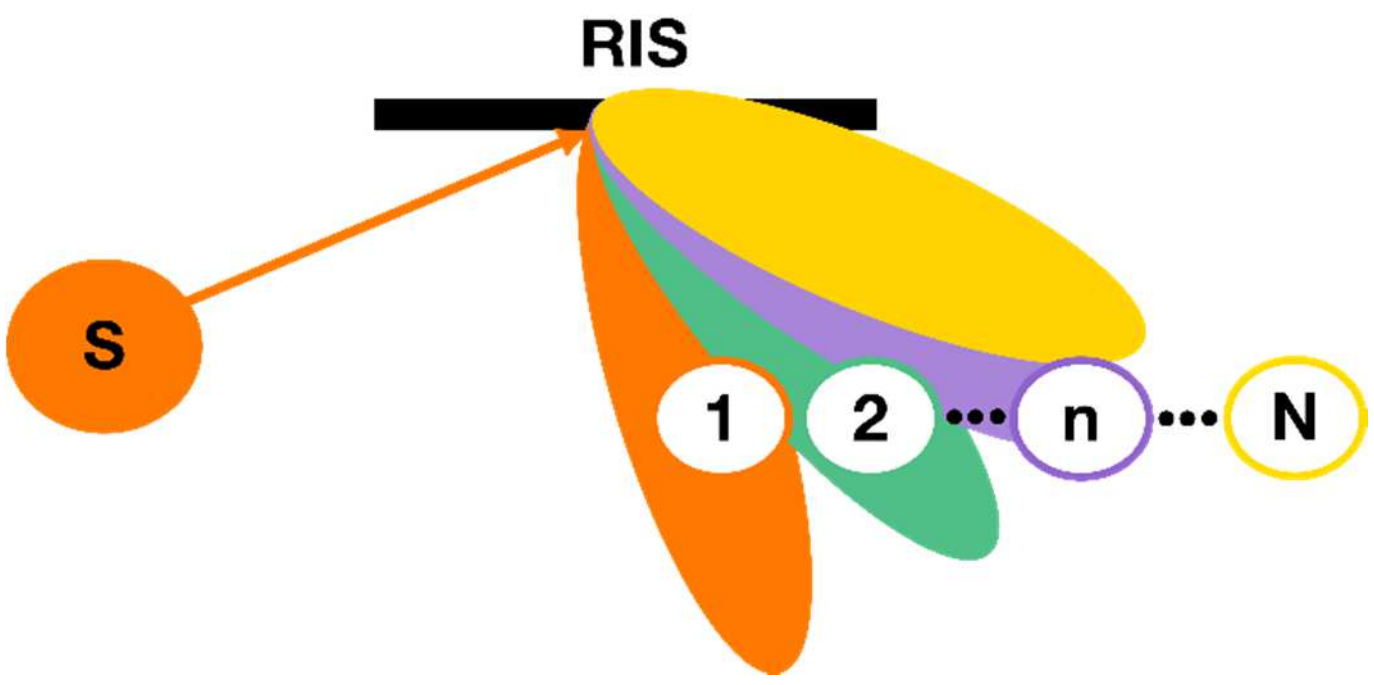

Fig. 2. Codebook principle.

We propose to improve $\Delta g$ with the RIS by using a codebook-based reflection beamforming approach. We first split the considered area (information desk) on which the tag and the reader are placed into $N$ predefined locations. For each location $n$, we compute the matrix of predefined reflection beamforming coefficients $\mathbf{b}^{(n)}$ that must be applied to the RIS, to create a reflected beam targeting the considered location $n$, as illustrated in Fig. 2. $\mathbf{b}^{(n)}$ is a diagonal matrix of unitary complex coefficients of size $M$. The codebook $\boldsymbol{B}$ is defined as the set of reflection beamforming matrices $\boldsymbol{B} = \{\mathbf{b}^{(1)} \dots \mathbf{b}^{(n)}, \dots, \mathbf{b}^{(N)}\}$, each matrix corresponds to a reflection beamformer that targets a predefined location. We build $\mathbf{u}$ as follows:

$$\mathbf{u} = \mathbf{b}e^{j\delta}, \quad (6)$$

where $\mathbf{b} \in \mathbf{B}$. With this notation, $\mathbf{b}_m$ is a cell-specific beam coefficient to be applied to the cell $m$ of the RIS, whereas $\delta \in \Delta$ is a beam phase-shift applied to all cells, that is chosen among a set of $P$ pre-defined beam phase-shifts $\Delta = \left\{0, \dots, \frac{2\pi}{P}, \dots, \frac{2\pi(P-1)}{P}\right\}$.

In order to compensate the phase shifts induced by the propagation channel (see equation (2) ) and therefore ensure a coherent focusing towards the target exact location $n$ (as illustrated in Fig. 3), the predefined reflected beamforming coefficient $\mathbf{b}_m{}^{(n)}$ of the m[th] cell of the RIS is computed from:

$$\mathbf{b}_m{}^{(n)} = e^{-j2\pi\frac{d_m^{S-Ris}+d_m^{Ris-n}}{\lambda}}, \quad (7)$$

Note that to compute the predefined coefficient $\mathbf{b}_m{}^{(n)}$ the location of the source and the RIS, and the propagation channels, must be known. The chosen spherical wave assumption is suitable for the considered scenario, where the source and the target are close to the RIS. Indeed, as illustrated in Fig. 3, in this case, the difference of beam coefficients between two successive cells is not constant as it takes into account the exact distances between all elements. Note that this simple computation method of reflecting beamforming coefficients does not take into account the mutual coupling between the cells of the RIS. We thus expect this method to be accurate enough to control for the direction of the main lobe of the reflected beam and less accurate for the control of the side lobes of the reflected beams.

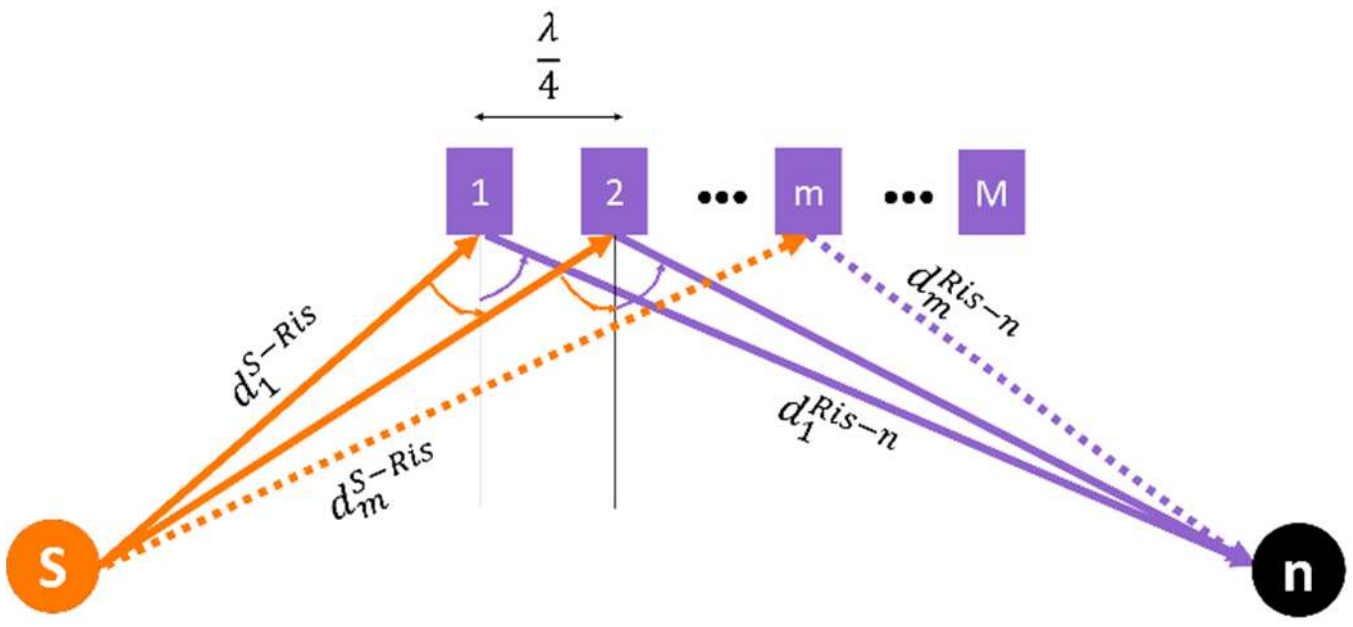

Fig. 3. Illustration of the model for the cells coherently targeting the location n, knowing the exact location of the source S.

The RIS tests $\mathbf{b}$ and $\delta$ pairs and the reader reports a quality feedback (a measured $\Delta g$ for instance). Together, the RIS and the reader find, through an exhaustive search, the $\mathbf{b}$ and $\delta$ pairs that maximise the performance. We expect at least two types of such pairs to exist such as $\Delta g^{Ris} > \Delta g^{ref}$.

In the first type of pairs, a "hot spot" beamformer $\mathbf{b}$ improves $\Delta g^{Ris}$ by boosting $h^{S-T-R,Ris}$ and reducing $h^{S-R,Ris} \approx 0$: thus, it creates a hot spot on the tag as in [4]. In this case, equation (4-b) becomes:

$$\Delta g^{Ris} \sim \left| \left| h^{S-R} + h^{S-T-R} + h^{S-T-R,Ris} \right| - \left| h^{S-R} \right| \right|. \quad (8)$$

In this case, $\delta$ must be chosen such that:

$$\arg(h^{S-R} + h^{S-T-R}) = \arg(h^{S-T-R,Ris})\ [2\pi], \quad (9)$$

which is equivalent to:

$$\begin{aligned}\delta &= \arg(h^{S-R} + h^{S-T-R}) \\ &\quad - \arg(\mathbf{h}^{S-Ris}\mathbf{b}\mathbf{h}^{Ris-T}h^{T-R})\ [2\pi].\end{aligned} \quad (10)$$

In the second type of pair, a "coherent spot" beamformer would attain the coherence condition as in [4]. More precisely, the coherence condition illustrated in Fig. 4 is obtained when the sum of the signals controlled by the RIS arrive at the reader location in phase with the sum of all the other signals. Such condition ensures that the two sums combine coherently and boost $\Delta g^{Ris}$. This coherence condition is expressed as follows:

$$\begin{aligned}&\arg(h^{S-R} + h^{S-T-R}) \\ &\quad = \arg(h^{S-T-R,Ris} + h^{S-R,Ris})\ [2\pi].\end{aligned} \quad (11)$$

We expect to find in the codebook a $\mathbf{b}$ that boosts $h^{S-T-R,Ris} + h^{S-R,Ris}$, for instance, by targeting at the same time the tag and the reader locations. For example, in some cases, this term can be boosted by two lobes of the same beam, pointing at the tag and the reader. Then, the following $\delta$ satisfies condition (10):

$$\begin{aligned}\delta &= \arg(h^{S-R} + h^{S-T-R}) \\ &\quad - \arg(\mathbf{h}^{S-Ris}\mathbf{b}\mathbf{h}^{Ris-R} + \mathbf{h}^{S-Ris}\mathbf{b}\mathbf{h}^{Ris-T}h^{T-R}).\end{aligned} \quad (12)$$

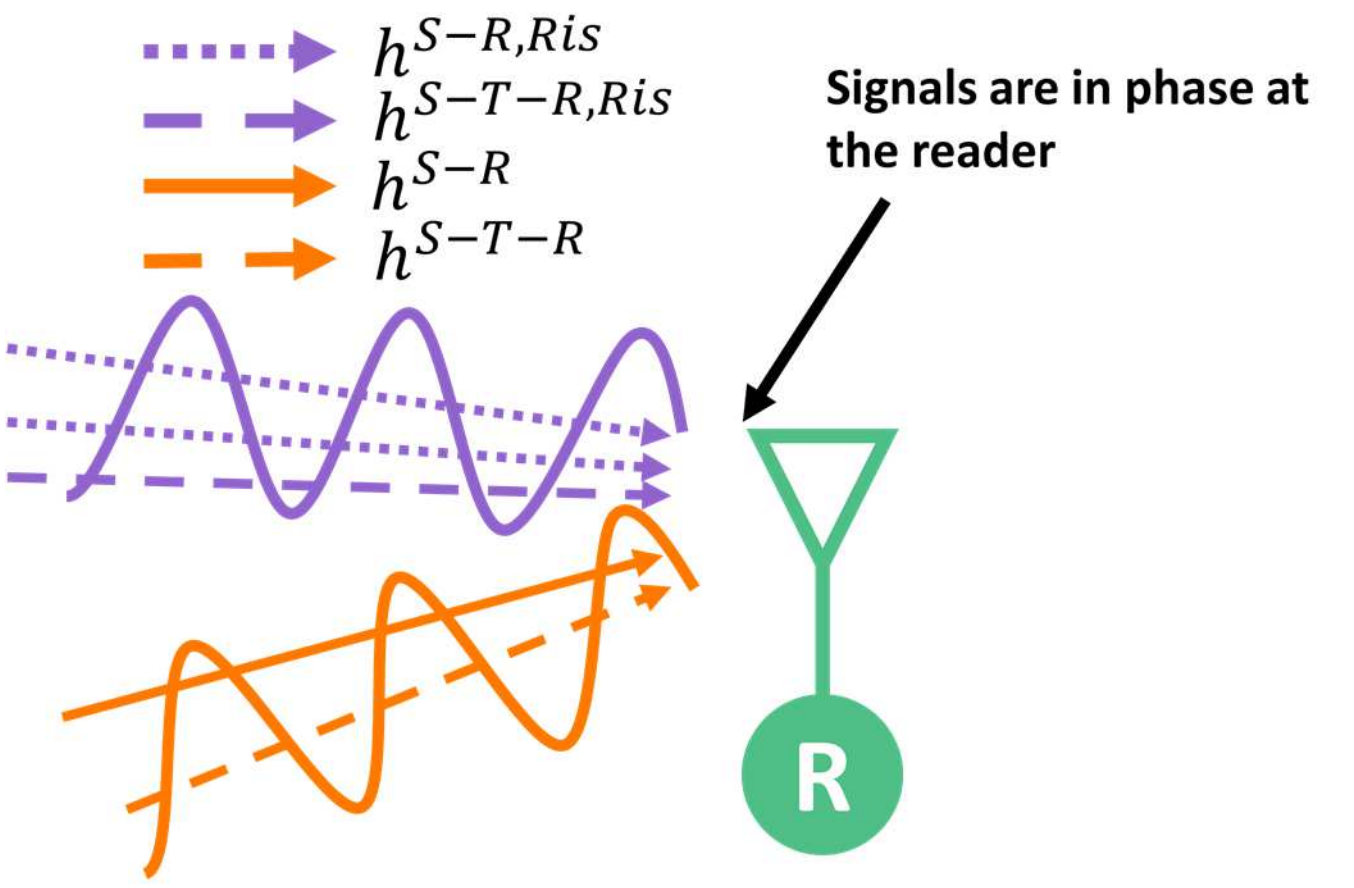

Fig. 4. Illustration of the coherent spots with the signals controlled by the RIS in phase with the direct signals (not passing through the RIS) at the reader.

## III. Experimental Setup

We conduct experimental measurements to validate the proposed RIS-assisted AmB communication system presented in Section II. As illustrated in Fig. 5, the measurements are set up on a desk, with all the devices (source, tag, reader and RIS) in the LOS of each other.

In order to evaluate the benefit of the RIS, two different configurations are tested and compared: with and without RIS.

We evaluate the performance of the AmB communication system in terms of BER for various **b** and $\delta$ pairs.

### *A. Experimental setup*

In the experimental setup, the source, the tag and the reader are all equipped with dipole antennas. The source generates an ambient signal with a 500kHz bandwidth at 5.2GHz. The reader receives the combined signals of the source, the RIS and the tag. A non-coherent energy detector is implemented in the reader to measure, in real time, the contrast $\Delta g$ defined in Section II. Both the source and the reader are controlled by an universal software radio peripheral (USRP) B210 on two separated channels. The software-defined radio "GNU Radio" is used to process the signals and acquire the data.

The tag switches between two states by disconnecting or connecting the two branches of its dipole as in [1]. It is electronically controlled thanks to a pin diode connected between the two branches of the antenna.

The coordinates of the location of each element of the system are denoted by $(x^E, y^E, z^E)$, where the element E can be the source (S), the tag (T), the reader (R) or the RIS. All the elements are located in the same horizontal desk, with $(z^S = z^T = z^R = z^{Ris})$ as vertical coordinates.

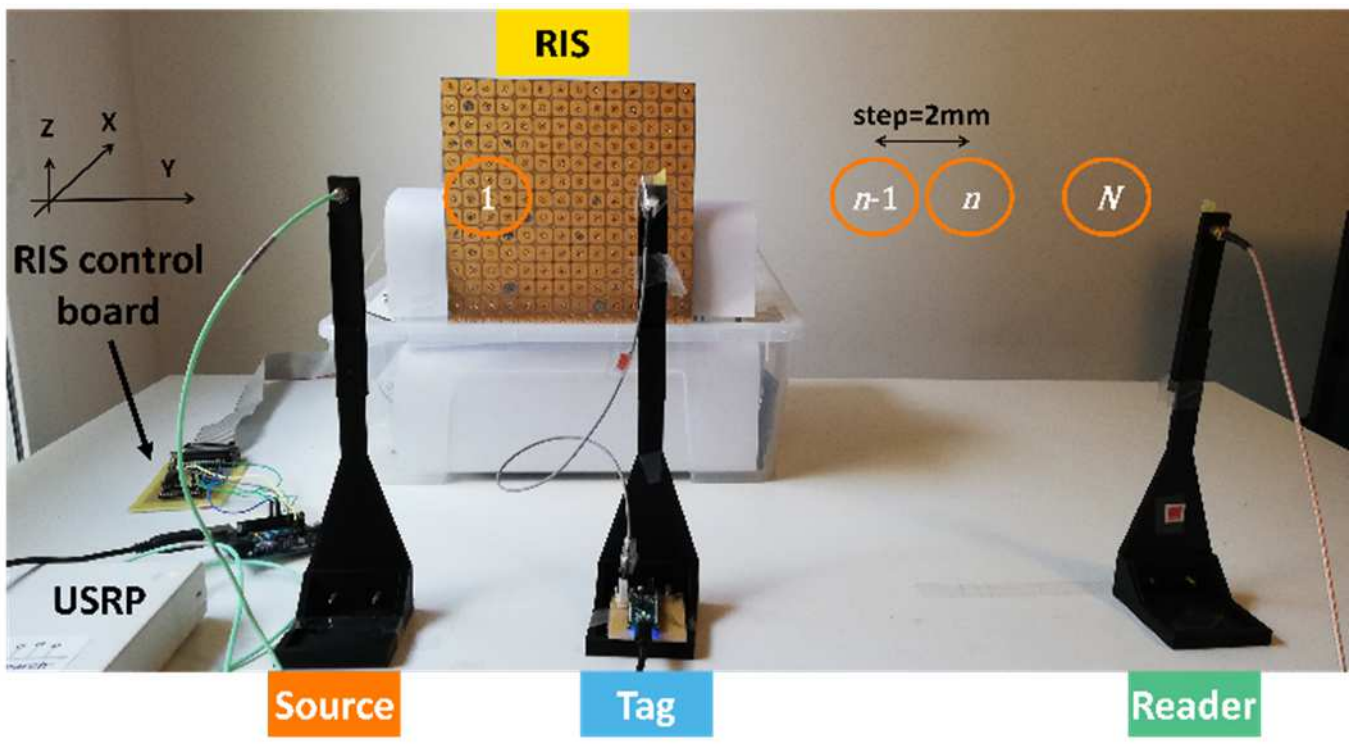

Fig. 5. Experimental setup.

### *B. RIS hardware prototype*

The RIS hardware prototype is a two-dimensional array composed of 14 rows and 14 columns, and is constituted, in total, by 196 cells. A single cell is illustrated in the Fig. 6. All cells have identical dimensions, i.e. 14×14mm and are designed for a frequency range between 5.15 to 5.75 GHz [15, 16]. The size of the cell is approximatively a quarter of the wavelength. The phase shift of each cell is controlled by applying a voltage between 0V and 5V, to a varactor implemented in each cell. The correlation between the phase shift and the voltage has been studied for this RIS and has been well characterized in [16]. In particular, the 14 columns of the RIS are controlled with a microcontroller connected to a digital-to-analog converter in order to provide the 0-5V control voltage to each column of cells.

In the experimental setup, the devices lie in the same horizontal plane, as shown in Fig. 5. Hence, we propose to exploit the symmetry of this configuration and limit the control of the RIS beamforming to the horizontal plane only. More precisely, the same beam coefficient is applied to each cell in the same column, and the 14 columns are controlled independently.

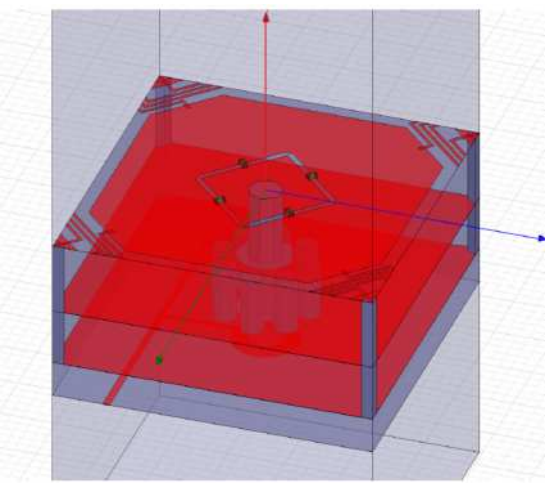

Fig. 6. Structure of the elementary cell of the RIS [15,16].

### *C. Codebook experimental implementation*

We first split the desk into $N$ pre-defined locations illustrated in Fig. 5. For each location $n$, we compute the corresponding codebook according to Section II and to the locations of the elements detailed in Table III. Note that as the inter-cell distance is a quarter of the wavelength, the mutual coupling between the RIS elements is not negligible and will impact the radiation pattern. However, the simple codebook computation method of Section II, i.e. without mutual coupling taken into account, is expected to be accurate enough for the control of the main direction and less accurate for the control of side lobes. The $N$ target locations are chosen along the y axis. As the elements are located on that same y axis, we vary the $y$ coordinate of the targets between $y = 0.1$m and $y = 0.4$m with a step of 2mm. This results in $N = 151$ different locations and a codebook of 151 different reflected beams.

Regarding the beam phase-shift $\delta$, we select $P = 18$ phase shift values between 0 and $2\pi$.

TABLE III. EXPERIMENTAL PARAMETERS

| *Parameters* | *Details* | *Value* | *Units* |
|---|---|---|---|
| $M$ | Number of independantly controlled RIS cells | 14 | |
| $N$ | Number of Beams | 151 | |
| $P$ | Number of Beam phase shifts | 18 | |
| $(x^S, y^S, z^S)$ | Location of the source | $(0, 0, 0)$ | m |
| $(x^{RIS}, y^{RIS}, z^{RIS})$ | Location of the center of the RIS | $(0, 0.1, 0)$ | m |
| $(x^T, y^T, z^T)$ | Location of the tag | $(0, 0.13, 0)$ | m |
| $(x^R, y^R, z^R)$ | Location of the reader | $(0, 0.57, 0)$ | m |

## IV. RESULTS

This section reports the simulation results obtained by using the simple model described in Section II and the experimental results obtained from the setup described in Section III.

The following parameters are set to the same values for the simulation and the experiments: the locations of the source, tag, reader and RIS, the carrier frequency, the pre-defined codebook and the predefined set of phase shifts (given in Section III-C).

For both the simulations and the experiments, all reflection beams of the codebook and all phase shifts are tested. For each pair of **b** and $\delta$, the contrast is assessed either by simulation or by direct experimental measurement at the reader side. In each case, the corresponding simulated BER and measured BER is derived. In addition, we conduct measurements in the absence of the RIS to provide a reference for comparison.

In Fig. 7-a) and b), we observe the simulated BER and measured BER, respectively, as a function of the index of the reflected beam applied **b** and the applied phase shift $\delta$. As expected, we observe, in Fig. 7, by simulations and experiments, that some pairs of reflected beam and beam phase-shift improve the performance: i.e., these correspond to the yellow pairs that minimize the BER.

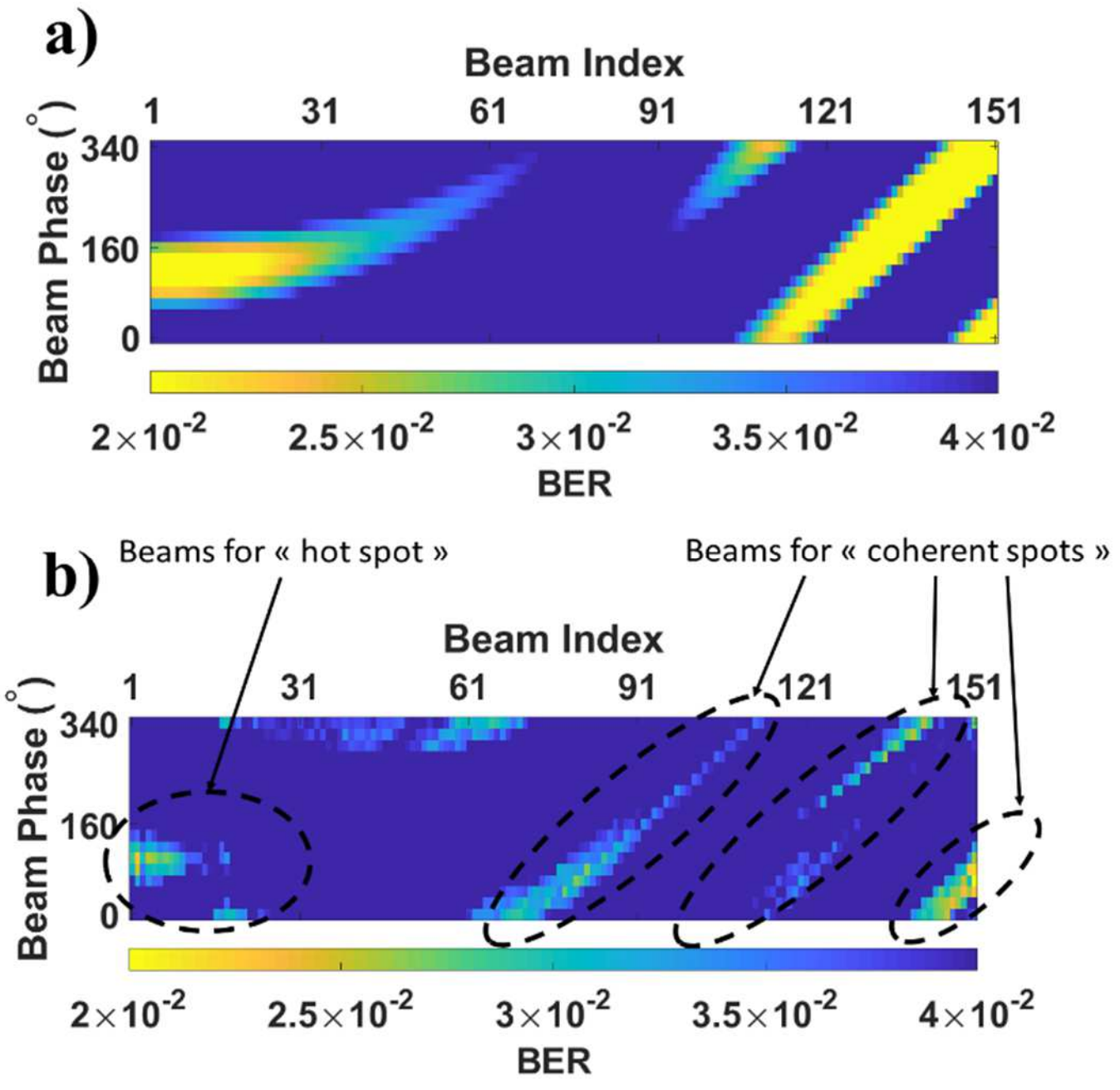

Fig. 7. Simulation (a) and experimental (b) results of the BER achievable depending on the beam of the codebook and the phase shift applied to the beam.

As expected, hot spots on the tag can be created with some reflected beams between 1 and 31 and with an optimised beam phase-shift fulfilling the condition in (8). One example of such reflected beam (the reflected beam 1) is illustrated in Fig. 8-a), where the simulated spatial map of the reflected signal strength is plotted in dB color scale. We observe, in this example, that the reflected beam 1 creates a hot spot on the tag thanks to its main lobe, whereas the reader receives almost no signal from the RIS. We also observe in Fig. 7.b) that, for the reflected beam 1, the condition (8) is approximatively obtained for $\delta = 140°$.

As expected, coherent spots can be obtained with several beams between 60 and 151 and with an optimized beam phase-shift verifying the condition in (8). Two examples of such beams (the reflected beams 68 and 145, respectively) are illustrated in Fig. 8-b) and c), with simulated maps of the reflected signal strength in dB color scale. As illustrated in Fig. 8-b) and c), respectively, the reflected beams 68 and 145, respectively, create coherent spots on both the tag and the reader, thanks to their main and side lobes. For both the reflected beams 68 and 145, we observe in Fig. 7.b) that the coherent condition in (10) is approximatively obtained for $\delta = 0°$.

We note that, in the simulated reflected beams, we did not take into account the mutual coupling. Hence, accurate information on the direction of the main lobe is obtained whereas the side lobes may be overestimated.

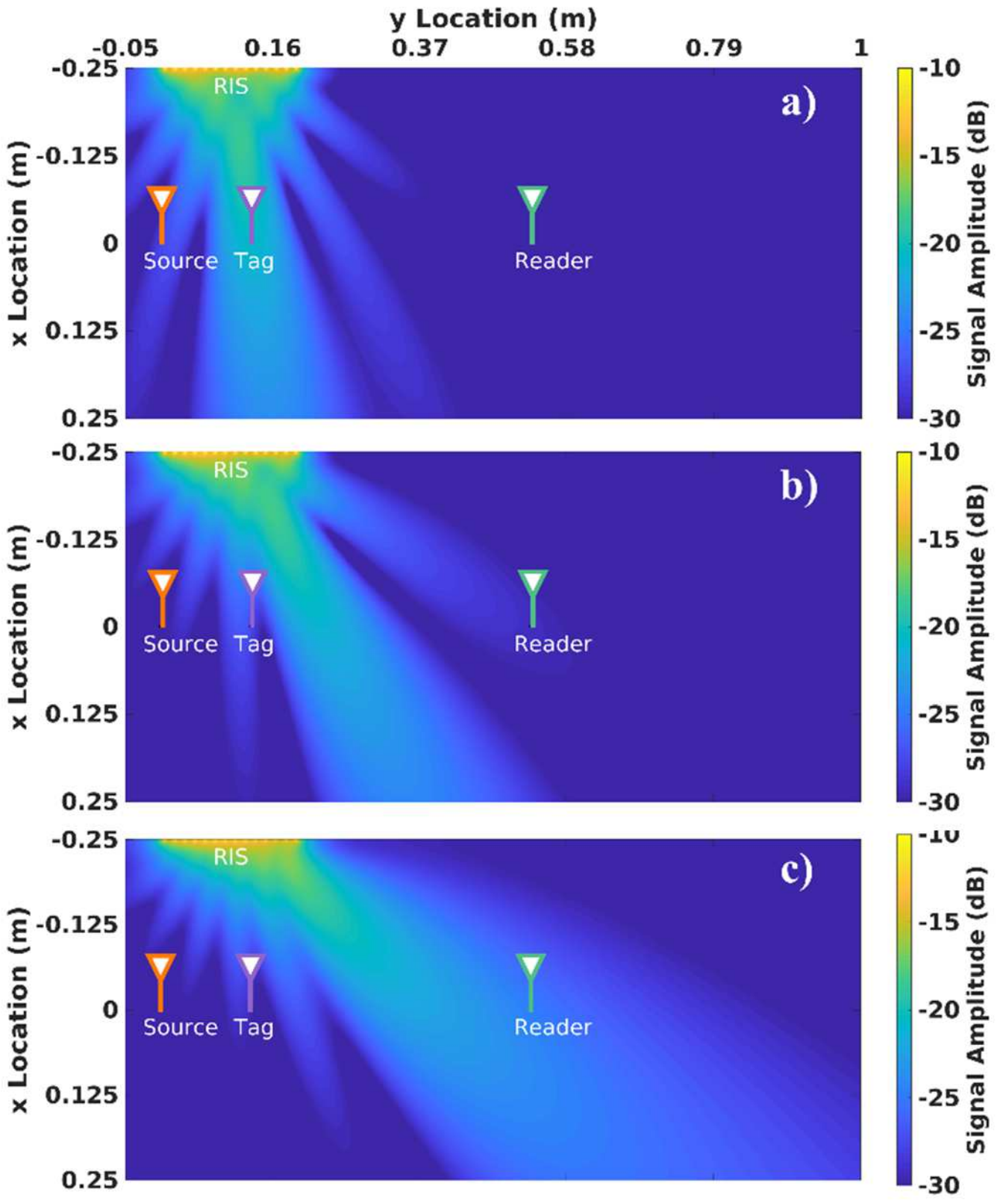

Fig. 8. Map illustrations of the beamformed reflected signal of the RIS for the reflection phase coefficient of beam 1 (a) and beam 68 (b) beam 145 (c) from the codebook.

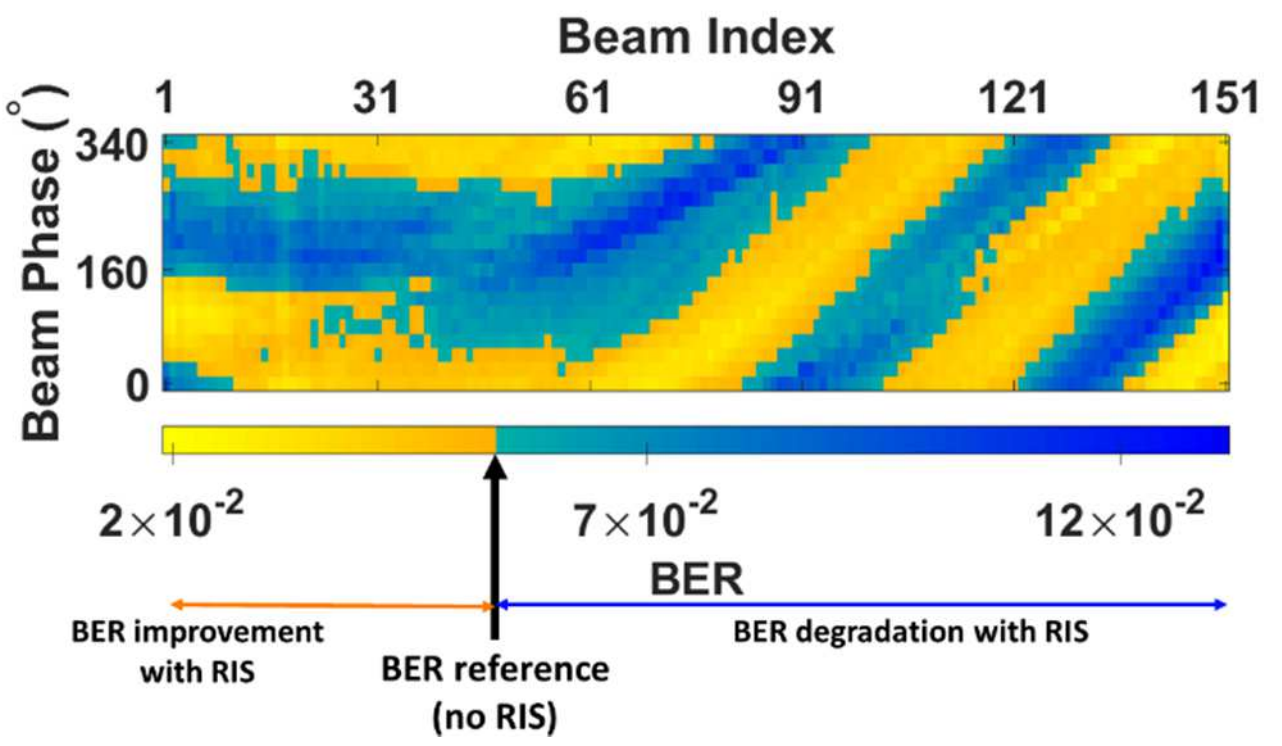

Fig. 9. Experimental results of the BER improvement/degradation due to the RIS depending on the beam of the codebook and the phase shift applied to the beam.

Finally, in Fig. 9, we provide a performance comparison of the AmB communication system with and without RIS. In the absence of the RIS, the system attains a reference BER value of $5.37\times10^{-2}$ that has been assessed through experimental measurements. In the presence of the RIS, the system attains a BER that is plotted in Fig. 9 as a function of the beam and phase shift pairs, in color scale. The chosen color scale highlights BER values exceeding the reference BER value (in blue) and BER values better than the reference BER value (in yellow). In yellow, we hence observe the pairs of $\mathbf{b}$ and $\delta$ that improve the BER compared to the configuration without RIS, whereas, in blue we have the RIS configurations that degrade the system performance.

These experimental results demonstrate that, assisted by an RIS that uses a codebook-based reflection beamforming scheme whose beams are chosen among a set of pre-defined beams for pre-defined locations and whose beam phase shifts are selected among a set of predefined values, the BER performance of an AmB communication system can be improved.

## V. Conclusion

In this paper, for the first time, we have demonstrated experimentally that reconfigurable intelligent surfaces can improve the performance of ambient backscatter systems. The proposed reconfigurable intelligent surface is able to realize pairs of a reflected beam (belonging to a codebook) and a beam phase-shift that improve the tag-to-reader bit-error-rate. Since the reconfigurable intelligent surface controls the reflected signal, a hot spot on the tag or coherent spots on the tag and the reader can be created to assist the communication between the tag and the reader. Future studies will evaluate more general codebooks that could be applied in different setups and also machine learning algorithms to optimize the phase reflection coefficients.

## Acknowledgment

This work has been partially supported by EU H2020 RISE-6G project.